\documentclass[letterpaper, 10 pt, conference]{ieeeconf}  

\IEEEoverridecommandlockouts                              

\title{\LARGE \bf
MemComputing vs. Quantum Computing: some analogies and major differences*
}

\author{Massimiliano Di Ventra, {\it Fellow, IEEE}$^{1}$ 
\thanks{*This work was supported by the National Science Foundation under Grant No.
	2034558.}
\thanks{$^{1}$Massimiliano Di Ventra is with the Department of Physics, University of California, San Diego La Jolla, CA 92093, USA 
        {\tt\small diventra@physics.ucsd.edu}}%
}

\IEEEoverridecommandlockouts
\pubid{\makebox[\columnwidth]{~
		\copyright~2022 IEEE \hfill} \hspace{\columnsep}\makebox[\columnwidth]{}} 
	
\begin{document}

\maketitle
\pagestyle{empty}

\begin{abstract}
Quantum computing employs some quantum phenomena to process information. It has been hailed as the future of computing but it is plagued by serious hurdles when it comes to its 
practical realization. MemComputing is a new paradigm that instead employs non-quantum dynamical systems and exploits  
time non-locality (memory) to compute. It can be efficiently emulated in software and its path towards hardware is more straightforward. I will discuss some analogies between these two computing paradigms, and the major differences 
that set them apart.


\end{abstract}

\section{INTRODUCTION}
There is a lot of discussion about ``unconventional computing''. What is meant by this is some type of computation that 
does not rely on the traditional Turing paradigm, as practically implemented with some architecture, such as the von Neumann one. However, before adding the adjective ``unconventional'' to computing, we should first answer the following: What does it mean to compute and what is its goal? 

It should be obvious, but it's worth stressing: the goal of a computing machine is to solve problems that are challenging for us, humans~\cite{Mybook}. Although we can define mathematically a plethora of ``computers'', most of them remain  
academic curiosities rather than useful tools in achieving such a goal.

Our modern computers (sometimes called ``classical''; see also Section~\ref{classical}) satisfy the above goal quite well: together with clever algorithms, they have been and still are extremely useful in solving a wide range of tasks that would be impossible for us, humans, to do by hand. Yet, even when the best algorithms are applied to hard instances of many 
combinatorial optimization problems, severe limitations emerge. The compute time for such instances increases exponentially when 
the size of the problem increases linearly. Such problems are the norm, not the exception, in both academic and 
industrial settings. 

Enter {\it Quantum Computing}~\cite{QI_bible}. In its theoretical underpinnings quantum computers (QCs) rely on qubits: quantum-mechanical systems 
with two states. To each qubit is associated a Hilbert space with basis states $|0\rangle$ and $|1\rangle$. 
A general state, $|\Psi\rangle$, in the Hilbert space of the qubit is a {\it linear 
	combination} of the type $|\Psi\rangle= 
\alpha_0|0\rangle+\alpha_1|1\rangle$, where $\alpha_0$ and $\alpha_1$ are some complex numbers such that $|\alpha_0|^2+|\alpha_1|^2=1$. 

Arguably the most famous application of QCs is prime factorization implemented with Shor's algorithm that if run on an {\it ideal} (i.e., decoherence-free) QC would factor numbers into primes in polynomial time~\cite{Shor-0}. The algorithm exploits two {\it physical} 
properties that are not available to our traditional computers: quantum entanglement and interference. 

Although limited in the size (and range) of problems they can actually solve, QCs have shown that physics-based 
approaches to computation provide advantages that are not easily obtained with traditional means. However, physical phenomena useful for computing  
are not just the domain of Quantum Mechanics. Other, equally valid phenomena can be exploited to solve hard combinatorial optimization 
problems.

Enter {\it MemComputing} (MC)~\cite{Mybook,diventra13a}. MC employs {\it time non-locality} (memory) to compute and relies on {\it non-quantum} dynamical systems as elementary building blocks~\cite{UMM}. The physics of MC is then fundamentally different from that of Quantum Computing. This translates into a 
completely different mathematical structure of MC with respect to Quantum Computing, with important consequences in both its software emulation and 
hardware realization. 

In this short paper, I will highlight some analogies between MemComputing and Quantum Computing, but most importantly their major differences. I will 
focus only on {\it digital} MC machines (DMMs), those mapping a finite string of symbols into a finite string of symbols~\cite{DMM2}. These are the ones that are easily scalable.     

\section{MemComputing vs. Quantum Computing}

\subsection{Collective computation and non-perturbative approaches}

The fundamental reason why some combinatorial optimization problems are challenging to solve is because they require {\it collective} 
assignments of variables in the problem specification~\cite{complexity-bible}. Take for instance the following Boolean formula relating the three logical variables $v_1$, $v_2$, $v_3$:
\begin{equation}\label{phix}
	\varphi({\bf v})=(\lnot v_{1}{\vee}v_{2}{\vee}v_{3}){\wedge}(\lnot
	v_{1}{\vee}\lnot v_{2}{\vee}v_{3}){\wedge} (v_{1}%
	{\vee}\lnot v_{2}{\vee}\lnot v_{3}),
\end{equation}
where ${\bf v}= (v_1,v_2,v_3)$, the symbols ${\vee}$ and ${\wedge}$ represent the logical OR and AND, 
respectively, while the symbol $\lnot$ represents negation, and the parentheses define a clause or constraint between variables. We want to know which logical assignment of ${\bf v}$ satisfies such as formula, that is when $\varphi({\bf v})=1$. 

Since the three variables appear in all three clauses, if we started from a random assignment (0 or 1) of the variables ${\bf v}$, we would need to check each clause separately to see if they are logically satisfied. Say, we start from the left of the formula~(\ref{phix}) and {\it sequentially} check if all clauses are satisfied. If not, we change the initial assignment of the variables and try again. This is easily done for such a simple formula. It becomes a considerable challenge if the number of clauses and variables grows in the thousands or even millions (which is standard in many academic and industrial applications).

Instead, it would be ideal to have a machine that can assign the correct value of such variables {\it collectively}, in the sense that it does not proceed sequentially but in parallel towards the solution, as if the machine were able to ``see'' the {\it global} structure 
of the problem (how the different variables ``interact'' in different clauses), not just its ``local'' features (the satisfaction of a single clause). 

Now, this type of global information is not so easy to extract with traditional algorithms, as clever as they can be. In fact, traditional algorithms can be classified as ``perturbative methods'' to computation: they change, following some rule-based strategy, the value of one or a few variables out of the many (large number of) variables in the problem specification. Instead, what is really needed is some sort of {\it non-perturbative 
approach}, where large numbers of variables (even comparable to the size of the problem) {\it simultaneously} change their values at different steps of the computation~\cite{Mybook}. 

The attribute ``non-perturbative" has a well defined meaning in physics. It means that the elementary units of
a physical system are {\it strongly coupled}, and we cannot describe their dynamics by separating the system into a non-interacting part, and then perturbing it by adding small interaction terms (``small'' compared to some energy scale characteristic of the system). Physical ideas seem to show up again. They suggest that we need to look for strongly coupled systems that showcase non-perturbative phenomena that we can exploit for 
computation. 

\subsection{Long-range order}
If we were to think of this problem in physical terms, with the variables representing actual physical quantities (e.g., voltages 
of an electrical circuit), and the logical gates as interactions between these quantities, we would immediately think of a machine that 
correlates these variables at all distances (wavelengths). Namely, we would think of a machine with {\it long-range order} (LRO)~\cite{Mybook}.

In the case of a QC we would invoke entanglement, since the latter does provide 
some type of collective behavior: a perturbation in (or measurement of) one of the system's parts, would immediately affect
other parts arbitrarily far away. This can be viewed as a type of long-range correlation or LRO. 

However, a quantum system in an entangled state has to be prepared experimentally in such as state (if it is not naturally in such a state). For instance, entanglement of qubits has to be realized experimentally at the beginning of the computation, and maintained 
during the whole duration of the computation for a QC to factor numbers~\cite{Shor-0}. And since entanglement is very sensitive to 
decoherence, maintaining it for long enough time for the computation to end is not an easy feat. 

\subsection{Dynamical LRO as an epiphenomenon of memory}

Instead, LRO can be found in a vast range of physical systems that do not showcase quantum dynamics. In fact, it may emerge as a 
natural byproduct of {\it time non-locality} even if the different parts of a physical system interact locally~\cite{Mybook}. To see this 
intuitively, suppose you have a collection of classical spins, $s_i$, on a lattice interacting locally, and their dynamics are described by the Hamiltonian:
\begin{equation}\label{eqn:Ising}
	H= -J\sum_{<ij>} s_i s_j , \;\; s_i \in \{-1, 1\},
\end{equation}
where the spin-spin interaction, with strength $J$, is short range (only nearest-neighbor spins interact). 

Suppose now that the spins have somehow memory of their past interaction, and the time it takes this memory to decay is 
much longer than the characteristic time of the spin-spin interaction: $t_c\sim 1/|J|$ (in appropriate units). This means that if a spin changes its value from, say, 1 to $-1$, or vice versa, its neighbors would respond to such a change. In turn, the latter ones will affect their 
neighbors, and so on, so that the interaction propagates through the lattice. However, if the spins have memory of their past dynamics, each one of them would effectively ``feel'' the interaction of all the other spins in the lattice through their time non-local response. In other words, time non-locality has naturally induced spatial non-locality even if the interactions are local~\cite{Mybook}. It is this property that DMMs exploit to solve hard combinatorial optimization problems efficiently~\cite{DMM2}. 

\subsection{Classical vs. non-quantum}\label{classical}
At this point it is worth making an important distinction. In some computer science literature the word ``classical'' is typically reserved for Turing machines, or our own traditional computers (the closest physical realization of deterministic Turing machines, but 
not exactly the same thing). This is what is meant when we read that quantum computers have reached (or not) ``quantum supremacy'' over ``classical computers''. 

MemComputing machines (in general, not just DMMs) are based on physical systems we would properly call ``classical''. However, 
they are not Turing machines and have nothing to do with our traditional computers. It is for this reason that I have used the 
word ``non-quantum'' rather than ``classical'' to describe their dynamics. 

\subsection{Phase space vs. Hilbert space}
That said, I can now move on to the mathematical structure of MemComputing vs. Quantum Computing which has important consequences on the applicability of 
these machines. As already mentioned, QCs are described by state vectors in a {\it Hilbert space}. This is a topological vector space~\cite{QI_bible}. 

Instead, MC machines, being non-quantum, are described by a state, ${\bf x}\equiv\{x_1,\dots,x_D\}$, with, say, $D$ components (degrees of freedom) in a $D$-dimensional topological manifold called the {\it phase space}. The dynamics of such machines are then described by ordinary differential equations (ODEs) of the type~\cite{Mybook}:
	\begin{equation}
	\dot {\bf x}(t) = F({\bf x}(t)); \;\;\;\;\;  {\bf x}(t=t_0)=  {\bf x}_0\label{ODE},
\end{equation}
where $F$ is a $D$-dimensional vector (the flow vector field), and the equations require a state assignment at an initial time, $t_0$. 

The distinction between the phase space of MC machines and the Hilbert space of a QC is substantial. The phase space is a topological manifold whose 
dimension grows {\it linearly} with the number of degrees of freedom. This means that if we map the variables of a combinatorial 
optimization problem into the degrees of freedom of a DMM, the phase space grows linearly with the problem size.

On the other hand, the Hilbert space of $N$ qubits is the tensor product of the individual qubits' Hilbert spaces. As such, it grows 
{\it exponentially} ($2^N$) with the number of qubits. 

Here then lies an incredible advantage of a DMM vs. a QC: a QC {\it cannot} be emulated efficiently on our traditional computers. It would require exponentially growing resources. A DMM instead {\it can} be simulated on such computers efficiently. This is because the random-access memory required to integrate numerically ODEs scales linearly with the number of degrees of freedom, and, if chaos is not present (as in the DMMs~\cite{Mybook}), the numerical stability of the simulations can be also controlled with a polynomial overhead. 

Therefore, the performance of DMMs, their robustness against noise, etc., 
can be tested in software, even before a hardware realization of the same is done. Furthermore, the reliance on non-quantum dynamical systems with memory makes the hardware realization of DMMs much more straightforward than that  
of QCs. In fact, time non-locality can be emulated using active elements, such as transistors~\cite{Pershin2010}. Therefore, a full implementation 
of these machines using CMOS is feasible~\cite{DMM2}, and they can operate at room temperature, rather than at the cryogenic temperatures typical of QCs.  

MC then offers a workflow production similar to that of our modern computers: the chip design and performance are first simulated and checked numerically before the chips are sent to production. This saves tons of time and money in the process. No such advantage exists for QCs: to see their actual performance, they have to be built in hardware, and any issue that arises at that stage needs to be solved at that point. 

A wide range of problems in the combinatorial optimization class have been already solved using emulations of DMMs; see~\cite{Mybook} and references therein, and the case studies reported by the company MemComputing, Inc. (www.memcpu.com). These simulations have shown 
considerable advantages compared to state-of-the-art algorithms applied to the same problems. Since numerical noise is ``worse'' than 
physical noise (the former accumulates with integration time; the latter is typically local in space and time), the simulations also show 
the robustness of the solution search by DMMs with respect to noise. 

\subsection{Topological computing and its physical vacua}
The above robustness (whether physical or numerical) is due to the fact that DMMs employ topological objects to compute~\cite{topo}. These
are {\it instantons}, namely a family of topologically nontrivial deterministic trajectories smoothly connecting pairs of critical points---those values of ${\bf x}$ satisfying the condition $F({\bf x}(t))=0$ in Eq.~(\ref{ODE})---of increasing stability in the phase space~\cite{Mybook}. 

Instantons are also the type of {\it non-perturbative} phenomena (like quantum tunneling to which they are related) we were 
after to solve hard combinatorial optimization problems efficiently. This means that when an instanton occurs in phase space it can involve a large number of degrees of freedom (variables); in fact, even as large as the total number of variables. In other words, instantons realize {\it physically} the collective 
dynamics I mentioned above. 

Note also that instantons occur ``spontaneously'', in the sense that once the dynamics of a DMM are initiated 
the machine enters this instantonic phase on its own~\cite{Mybook}. This collective dynamical behavior does not need to be prepared experimentally 
at the beginning of the computation, like entanglement of a QC. It occurs during time evolution. 

In addition, the transition amplitudes between any two critical points are {\it topological invariants} on instantons, even in the presence of noise~\cite{DMMtopo}. This means that they cannot change without changing 
the topology of phase space. Importantly, the number of instantonic steps required to reach the solution of the problem at hand can be easily counted~\cite{DMMtopo}: it grows {\it polynomially} with the size of the problem to solve, even in the presence of moderate noise~\cite{Mybook}. 

Another type of topological computing has been suggested for QCs~\cite{TQC1}, but has yet to be realized in practice. It would rely on some strongly correlated electron systems to compute protected against the unavoidable decoherence. The mathematical framework to describe this
form of computation is a Schwarz-type {\it topological field
theory} (TFT)~\cite{book}. Instead, to describe the dynamics of the non-quantum dynamical systems representing DMMs a Witten-type TFT has been employed~\cite{topo,DMMtopo}. It is then interesting to see that a TFT underlies the description of the physical vacua of 
topological computation, whether quantum or not. Research on this analogy and its consequences in computing would be thus desirable. 

\subsection{Linear or non-linear machines?}
Despite the above analogy, another major difference distinguishes DMMs from QCs (topological or not). DMMs rely on {\it non-linear} phenomena, 
like instantons, to compute. On the other hand QCs are fundamentally linear machines, if we do not consider the measurement process,
and possible coupling with the environment (which leads to decoherence). QCs manipulate
state vectors in a linear vector space (a Hilbert space) using linear operators (quantum gates) on that space. As I mentioned in the 
Introduction, the price that needs to be paid for this linearity is the exponential growth of the Hilbert space dimension with the number 
of qubits employed. 

On the other hand, DMMs take advantage of non-linearities (instantons) to compute efficiently with the dimension of the phase space growing linearly with the number of degrees of freedom (size of the problem to solve). In other words, DMMs have traded 
homogeneity and superposition, typical of linear systems, with non-linear dynamics, to gain linearity in the scaling of the phase space. As I have discussed above, this trade-off favors MemComputing over Quantum Computing. 

\subsection{Deterministic vs. probabilistic computing}
Let me finally mention that DMMs (MC in general) are {\it deterministic} dynamical systems. Noise is a nuisance, not a fundamental 
tool to compute. QCs instead are {\it probabilistic} machines: the calculation of a given problem needs to be repeated many times on
an equally prepared system, so that an average of the results can be collected. This is an intrinsic and unavoidable feature of Quantum
Mechanics. 

Although the act of repeating a measurement many times (for the benefit of obtaining an average quantity with a certain level of confidence) is not {\it per se} a show stopper, it is definitely not advantageous if the same problem can be solved with just one attempt. A deterministic machine holds then a considerable advantage compared to a probabilistic one. This advantage is even more pronounced if noise, in the form of say, temperature fluctuations, is used as the fundamental computing tool: such a probabilistic machine would need to 
navigate an exponentially growing {\it state space} with consequent detriment to its scalability as a function of problem size. 

\section{CONCLUSIONS}
In conclusion, I have discussed a few analogies and the major differences between MemComputing and Quantum Computing. These two 
physics-based paradigms of computation are fundamentally different in both the physics they employ and their mathematical description. These differences have profound consequences on their applicability to hard combinatorial optimization problems. MemComputing machines 
can be efficiently emulated in software (showing already substantial advantages compared to traditional algorithms), while QCs need to 
be built in hardware to realize their potential. Irrespective of these differences, the two paradigms can provide fruitful cross-pollination of ideas with possible benefits to both. Research in this direction would then be beneficial. 


\end{document}